\documentclass[preprint,amsmath,amssymb,aps,showkeys,showpacs]{revtex4}
\usepackage[english]{babel}
\usepackage{graphicx}
\usepackage{graphics}
\usepackage{amsmath}
\usepackage{dcolumn}
\usepackage{amssymb}
\usepackage{bm}

\begin{document}
\title{On the missing magnetic flux and topological effects of a screw dislocation on a charged particle in an inhomogeneous magnetic field}
\author{K. Bakke}
\email{kbakke@fisica.ufpb.br}
\affiliation{Departamento de F\'isica, Universidade Federal da Para\'iba, Caixa Postal 5008, 58051-900, Jo\~ao Pessoa, PB, Brazil.}

\author{C. Furtado}
\email{furtado@fisica.ufpb.br}
\affiliation{Departamento de F\'isica, Universidade Federal da Para\'iba, Caixa Postal 5008, 58051-900, Jo\~ao Pessoa, PB, Brazil.}

\begin{abstract}

We study the interaction of an electron/hole with inhomogeneous magnetic field in the presence of a screw dislocation. We consider the internal structure of the defect, i.e., we consider the core that gives rise to a finite size to the defect. In addition, we assume that this core determines a forbidden region for the electron/hole. Then, we solve the Schr\"odinger equation and show that an Aharonov-Bohm-type effect arises from the influence of the topological defect and the missing magnetic flux on the eigenvalues of energy.

\end{abstract}


\maketitle

\section{Introduction}

The study of the influence of topology and geometry on quantum particle dynamics has the Aharonov-Bohm effect \cite{ab} as an important system for the quantum physics world. In the fifties, Aharonov and Bohm \cite{ab} demonstrated the quantum physical effect of the electromagnetic vector potential. They investigated the quantum dynamics of an electron around a long, thin solenoid. They found that the electron wave function acquires a quantum phase, which has been called as the Aharonov-Bohm phase. What was observed after this quantum phenomenon is manifested in several areas of physics \cite{pesk,hagen} and has important experimental applications in mesoscopic systems \cite{loss,bagache}. In mesoscopic systems, the presence of conductance oscillations was experimentally observed as a function of the magnetic flux. This effect has been observed for Aharonov-Bohm experiment in quantum dots \cite{yac,schus, heib, heib2,van}.

In recent years, there has been a great interest in two dimensional electron gases submitted to inhomogeneous magnetic fields \cite{noga1,xu,noga2} due to a variety of physical systems that can be studied and several physical phenomena, such as the quantum Hall effect and superconductivity \cite{noga1}. Several authors have investigated experimental set-ups with the aim of producing non-uniform magnetic fields by using patterned gates made out of superconductors \cite{bend,geim} and ferromagnets \cite{weiss,noga3,lange}. In Ref.\cite{lind}, an experimental arrangement that produces inhomogeneous magnetic fields was obtained by using a thin film of type II superconductors. In the theoretical point view, several proposals of an inhomogeneous magnetic field configuration have been studied. In Refs.\cite{down1,down2,down3,eshg, sourr,roy,wang}, massless fermions in the presence of inhomogeneous magnetic field have been studied with interest in quantum dots, quantum rings and other nanostructures in graphene physics. 

In Ref. \cite{vagner}, the authors have proposed a class of topological phenomena in absence of an external magnetic field in a mesoscopic quantum ring. In Ref \cite{magdot3}, the quantum dynamics of an electron in magnetic quantum dot was investigated. The edge states in the magnetic quantum dot show that the electrons are apparently confined to the plane. Besides, the eigenvalues of energy are a function of a parameter that depends on magnetic field. This parameter is called as the missing magnetic flux and it is defined by $\varsigma=\frac{B_{0}\,\pi\,r_{0}^{2}}{\phi_{0}}$, where $\phi_{0}=\frac{h}{e}$ is the flux quanta, $B_{0}$ is the intensity of the magnetic field and $r_{0}$ is the radius of the quantum dot. The parameter $\varsigma$ represents the number of missing magnetic flux within the quantum dot.

In this work, we bring a discussion about the topological effects of a thick screw dislocation on the interaction of a point charge (electron or hole) with an inhomogeneous magnetic field. We use the geometric theory of defects \cite{kat,extra1,furtado3,extra2} to describe the geometry of an elastic medium with a screw dislocation. Then, the magnetic field does not fill the region of radius $r_{0}$, where $r_{0}$ is the radius of the topological defect core. In addition, we assume that this region of radius $r_{0}$ is a forbidden region for the point charge. Thereby, we show that bound state solutions to the Schr\"odinger equation can be achieved, where the spectrum of energy depends on the missing magnetic flux. We also discuss the Aharonov-Bohm-type effect \cite{ab,pesk} and persistent currents \cite{by} associated with the missing magnetic flux.

The structure of this paper is: in section II, we start by describing the inhomogeneous magnetic field and by introducing the line element that describes the screw dislocation. Then, we analyse the interaction of a point charge with this inhomogeneous magnetic field in the elastic medium with a screw dislocation by assuming that the region where the magnetic field is absent is forbidden for the point charge. We show that analytical solutions to the Schr\"odinger equation can be obtained. Furthermore, we obtain the allowed energies of the system in a particular case, and thus, we discuss Aharonov-Bohm-type effects \cite{ab,pesk}; in section III, we discuss the appearance of persistent currents; in section IV, we present our conclusions.

\section{Topological effects}

Let us study the interaction of a point charge (electron or hole) with an inhomogeneous magnetic field in an elastic medium with a screw dislocation. This magnetic field fills the whole space except of a region of radius $r_{0}$ (it can be seen as a region inside a cylinder of radius $r_{0}$ that characterizes the thick defect, where $r_{0}$ is radius of the core of the topological defect). Hence, the magnetic field is given by $B=0$ for $r\,<\,r_{0}$ and $\vec{B}=B_{0}\,\hat{z}$ for $r\,>\,r_{0}$, $B_{0}>0$ is a constant. As shown in Refs. \cite{magdot,magdot2}, the vector potential is null when $r\,<\,r_{0}$ and for $r\,>\,r_{0}$ is given by
\begin{eqnarray}
\vec{A}\left(r\right)=\frac{B_{0}\left(r^{2}-r_{0}^{2}\right)}{2\,r}\,\hat{\varphi}.
\label{1.1}
\end{eqnarray}

In addition, let us consider an elastic medium with a screw dislocation. This topological defect corresponds to the distortion of a circular curve into a vertical spiral \cite{val,put}. Hence, from the Katanaev-Volovich approach \cite{kat}, this screw dislocation can be described by the line element:
\begin{eqnarray}
ds^{2}=dr^{2}+r^{2}d\varphi^{2}+\left(dz+\beta\,d\varphi\right)^{2}.
\label{1.2}
\end{eqnarray}
The parameter $\beta$ is a constant that characterizes the dislocation. Its relation to the Burgers vector $\vec{b}$ is given by $\beta=b/2\pi$. In addition, the Burgers vector of the distortion of a circular curve into a vertical spiral is perpendicular to the plane $z=0$.

Hence, the time-independent Schr\"odinger equation that describes the interaction of the point charge with the magnetic field, in an elastic medium with a topological defect, is written in the following form \cite{fur,furtado3,fur6,l1} (we shall use the units $\hbar=1$ and $c=1$):
\begin{eqnarray}
E\psi=-\frac{1}{2m}\frac{1}{\sqrt{g}}\,\left(\partial_{k}-iq\,A_{k}\right)\left[\sqrt{g}\,g^{kj}\left(\partial_{j}-iq\,A_{j}\right)\right]\psi,
\label{1.3}
\end{eqnarray} 
where $g_{ij}$ is the metric tensor, $g^{ij}$ is the inverse of $g_{ij}$, and thus, $g=\mathrm{det}\left|g_{ij}\right|$. Besides, $q$ corresponds to the electric change and $A_{k}$ is a covariant component of the electromagnetic four-vector potential $A_{\mu}=\left(A_{0},\,A_{k}\right)$. With the screw dislocation (\ref{1.2}) and the vector potential (\ref{1.1}), hence, after some calculations, the time-independent Schr\"odinger equation becomes
\begin{eqnarray}
E\psi&=&-\frac{1}{2m}\left[\frac{\partial^{2}}{\partial r^{2}}+\frac{1}{r}\frac{\partial}{\partial r}+\frac{1}{r^{2}}\left(\frac{\partial}{\partial\varphi}-\beta\frac{\partial}{\partial z}\right)^{2}+\frac{\partial^{2}}{\partial z^{2}}\right]\psi+\frac{i\,q\,B_{0}}{m\,r}\left(\frac{r^{2}-r_{0}^{2}}{2\,r}\right)\left(\frac{\partial}{\partial\varphi}-\beta\frac{\partial}{\partial z}\right)\psi\nonumber\\
&+&\frac{q^{2}B_{0}^{2}}{8m}\frac{\left(r^{2}-r_{0}^{2}\right)^{2}}{r^{2}}\psi
\label{1.4}
\end{eqnarray}

Since this system has the cylindrical symmetry, thus, the solution to Eq. (\ref{1.4}) can be written in terms of the eigenvalues of the $z$-components of the angular momentum and the linear momentum operators: $\psi\left(r,\,\varphi,\,z\right)=e^{il\varphi+ikz}\,u\left(r\right)$, where $k$ is a constant and $l=0,\pm1,\pm2,\pm3,\pm4\ldots$. With this solution, we obtain the following radial equation: 
\begin{eqnarray}
u''+\frac{1}{r}\,u'-\frac{\gamma^{2}}{r^{2}}\,u-\frac{m^{2}\omega^{2}}{4}\,r^{2}\,u+\tau\,u=0,
\label{1.5}
\end{eqnarray}
where the parameters $\omega$, $\gamma$ and $\tau$ are defined as follows:
\begin{eqnarray}
\omega&=&\frac{q\,B_{0}}{m};\nonumber\\
\gamma&=&l-\beta\,k+\frac{q\,\varsigma}{2\pi};\label{1.6}\\
\tau&=&2mE+m\omega\,\gamma-k^{2}.\nonumber
\end{eqnarray}
Observe that the parameter $\varsigma$ that appears in Eq. (\ref{1.6}) is the missing magnetic flux quanta \cite{magdot3}, i.e., it is the magnetic flux that is missed from the uniform magnetic field inside the region of radius $r_{0}$: $\varsigma=B_{0}\,\pi\,r_{0}^{2}$. It can also viewed through the vector potential (\ref{1.1}). By taking the term proportional to $r_{0}^{2}$ of the vector potential (\ref{1.1}), we obtain
\begin{eqnarray}
\varsigma=\oint\,\frac{B_{0}\,r_{0}^{2}}{2r}\,\hat{\varphi}\cdot\,d\vec{r}=\,B_{0}\,r_{0}^{2}\,\pi.
\label{1.6a}
\end{eqnarray}
As pointed out in Ref. \cite{magdot}, when the point charge interacts with the inhomogeneous magnetic field described in Eq. (\ref{1.1}), the system can be considered as the inverse of the Aharonov-Bohm proposal \cite{ab,pesk}. This means that the missing magnetic flux inside the region $r\,<\,r_{0}$ can influence the interaction of the point charge with the magnetic field in the region $r\,>\,r_{0}$. As a result, the missing magnetic flux $\varsigma$ yields the effective angular momentum $\gamma$. An interesting point to be observed in Eq. (\ref{1.6a}) is that there is no influence of the topology of the screw dislocation on the missing magnetic flux (\ref{1.6a}).

Next, let us perform the change of variables $y=\frac{m\omega}{2}\,r^{2}$, and then, Eq. (\ref{1.5}) becomes:
\begin{eqnarray}
y\,u''+u'-\frac{\gamma^{2}}{4y}\,u-\frac{y}{4}\,u+\frac{\tau}{2m\omega}\,u=0.
\label{1.7}
\end{eqnarray}

By analysing the behaviour of Eq. (\ref{1.7}) as $y\rightarrow\infty$, we can write the solution to Eq. (\ref{1.7}) in terms of the Whittaker function $W_{\tau,\,\,\left|\gamma\right|/2}\left(y\right)$, i.e.,
\begin{eqnarray}
u\left(y\right)=\frac{1}{\sqrt{y}}\,W_{\tau,\,\frac{\left|\gamma\right|}{2}}\left(y\right).
\label{1,7a}
\end{eqnarray} 
Note that the Whittaker function $W_{\mu,\,\frac{\left|\gamma\right|}{2}}\left(y\right)$ is regular at $y\rightarrow\infty$ \cite{abra}. Let us use the relation of $W_{\kappa,\,\nu}\left(y\right)$ to the confluent hypergeometric function of the second kind $U\left(a,\,b;\,y\right)$ \cite{abra,arf}:
\begin{eqnarray}
W_{\kappa,\,\nu}\left(y\right)=e^{-y/2}\,y^{\frac{1}{2}+\nu}\,U\left(\frac{1}{2}+\nu-\kappa,\,1+2\nu;\,y\right),
\label{1.7b}
\end{eqnarray}
where $U\left(a,\,b;\,y\right)$ is also regular at $y\rightarrow\infty$ \cite{abra}. The parameters $\kappa$ and $\nu$ that appear in the relation (\ref{1.7b}) are given by $\kappa=\frac{b}{2}-a=\frac{\tau}{2m\omega}$ and $\nu=\frac{b}{2}-\frac{1}{2}=\left|\gamma\right|/2$, thus, we have that $U\left(\frac{\left|\gamma\right|}{2}+\frac{1}{2}-\frac{\tau}{2m\omega},\,\left|\gamma\right|+1;\,y\right)$. Let us work with the confluent hypergeometric function of the second kind, hence, the solution to Eq. (\ref{1.7}), regular at $y\rightarrow\infty$, is given as follows: 
\begin{eqnarray}
u\left(y\right)=e^{-\frac{y}{2}}\,y^{\left|\gamma\right|/2}\,U\left(\frac{\left|\gamma\right|}{2}+\frac{1}{2}-\frac{\tau}{2m\omega},\,\left|\gamma\right|+1;\,y\right).
\label{1.8}
\end{eqnarray}

From now on, let us assume that $r=r_{0}$ is a cut-off point. This imposes that the wave function $u\left(r\right)=0$ at $r=r_{0}$. In terms of the dimensionless parameter $y$, when $r=r_{0}$, then, $y_{0}=\frac{m\omega\,r^{2}_{0}}{2}$. Thus, we have the boundary condition:
\begin{eqnarray}
u\left(y_{0}=\frac{m\omega r^{2}_{0}}{2}\right)=0.
\label{1.11}
\end{eqnarray}

From this perspective, we have from the boundary condition (\ref{1.11}):
\begin{eqnarray}
u\left(y_{0}\right)\Rightarrow U\left(\frac{\left|\gamma\right|}{2}+\frac{1}{2}-\frac{\tau}{2m\omega},\,\left|\gamma\right|+1;\,y_{0}\right)=0.
\label{1.13}
\end{eqnarray}

Let us discuss a particular case of the function $U\left(a,\,b;\,y_{0}\right)$ given when we have a fixed value $y_{0}$. With the fixed $y_{0}$ and a fixed value for the parameter $b=\left|\gamma\right|+1$, then, by considering $\frac{\tau}{2m\omega}\gg1$, thus, the parameter $a=\frac{\left|\gamma\right|}{2}+\frac{1}{2}-\frac{\tau}{2m\omega}$ can be considered to be large. Then, the function $U\left(a,\,b;\,y_{0}\right)$ can be written in the form \cite{abra}: 
\begin{eqnarray}
U\left(a,\,b;\,y_{0}\right)\propto\cos\left(\sqrt{2by_{0}-4ay_{0}}-\frac{b\pi}{2}+a\pi+\frac{\pi}{4}\right).
\label{1.14}
\end{eqnarray}

Therefore, by using Eq. (\ref{1.6}) after substituting Eq. (\ref{1.14}) into Eq. (\ref{1.13}) , we obtain the energy levels:
\begin{eqnarray}
E_{n,\,l}=-\omega\left[n+\frac{1}{2}\left(l-\beta\,k+\frac{q\,\varsigma}{2\pi}\right)+\frac{1}{4}\right]+\frac{q\,\varsigma}{\pi^{3}}\,\omega\left[1\pm\sqrt{1-\frac{\pi^{3}}{2\,q\,\varsigma}\left(4n+1\right)}\right]+\frac{k^{2}}{2m},
\label{1.15}
\end{eqnarray}
where $n=0,1,2,3,\ldots$ is the radial quantum number.

Hence, the energy levels (\ref{1.15}) arises from the interaction of a point charge with an inhomogeneous magnetic field in an elastic medium with a screw dislocation under the influence of a cut-off point. These bound states are achieved in the region $r>r_{0}$. We have that the topology of the screw dislocation influences the energy levels (\ref{1.15}). It can be viewed through the presence of the parameter $\beta$ in the energy levels (\ref{1.5}). Therefore, the effects of the topology of the screw dislocation yields a contribution to the angular momentum quantum number that gives rise to the effective angular momentum $\gamma$. Since no interaction between the point charge and the defect exists, this influence of the topological defect on the energy levels gives rise to an Aharonov-Bohm-type effect for bound states \cite{pesk,fur,fur2,fur3,vb,vb2,vb3}. By taking $\beta=0$, we obtain the energies levels in the absence of the screw dislocation.   

Besides, the energy levels (\ref{1.15}) depend on the missing magnetic flux $\varsigma$, which means that the missing magnetic flux determines the degeneracy of the energy levels. Therefore, the missing magnetic flux inside the region $r\,<\,r_{0}$ can influence the interaction of the point charge with the magnetic field in the region $r\,>\,r_{0}$. This agrees with Refs. \cite{magdot,magdot2,magdot3}, where the missing magnetic flux modifies the degeneracy of the Landau levels. Furthermore, according to Ref. \cite{magdot}, this system can be considered as the inverse of the Aharonov-Bohm proposal \cite{ab}. Therefore, the dependence of the energy levels (\ref{1.15}) on the missing magnetic flux $\varsigma$ can be considered as an analogue of the Aharonov-Bohm effect for bound states \cite{ab,pesk}, in the sense that it can be considered as an inverse of the proposal of the Aharonov-Bohm effect for bound states \cite{pesk}. In this way, we call this quantum effect as the Aharonov-Bohm effect for missing magnetic flux.

Another aspect of the energy levels (\ref{1.15}) is the fact that the radial quantum number is restricted to
\begin{eqnarray}
n\,<\,\frac{\left|q\right|\varsigma}{2\pi^{3}}-\frac{1}{4}.
\label{1.16}
\end{eqnarray}
Otherwise, we would have an imaginary term in the spectrum of energy. Therefore, the upper limit of the radial quantum number is determined by the missing magnetic flux. Finally, the last term of Eq. (\ref{1.15}) corresponds to the translational kinetic energy that arises from the free motion of the particle in the $z$-direction.

\section{persistent currents}

The persistent current is an interesting phenomenon observed in quantum ring systems and has been studied in the theoretical and experimental point of views \cite{per4,per3,per5,per6,per7}. The persistent current can be simply physically interpreted as that the magnetic flux enclosed by the quantum ring will introduce an asymmetry between electrons with clockwise and counterclockwise momenta, and thus, lead to a thermodynamic state with a persistent current \cite{per8}. As we have seen in Eq. (\ref{1.15}), the energy levels depend on the missing magnetic flux. Therefore, the missing magnetic flux $\varsigma$ influences the interaction of the point charge with the magnetic field in the region $r\,>\,r_{0}$. Moreover, this dependence of the energy levels on the missing magnetic flux can be viewed as the inverse of the proposal of the Aharonov-Bohm effect for bound states \cite{pesk}, i.e., it can be viewed as the Aharonov-Bohm effect for missing magnetic flux.

Another interesting point to be raised from the dependence of the energy levels on the missing magnetic flux is the possibility of having an analogue of the persistent currents \cite{by}. This is justified by the fact that the missing magnetic flux $\varsigma$ is determined by the intensity of the magnetic field $B_{0}$. In this way, if we increase or decrease $B_{0}$, i.e., if we vary the intensity of the magnetic field in the region $r\,>\,r_{0}$, thus, the missing magnetic flux varies inside the region $r\,<\,r_{0}$. In this sense, the definition of persistent currents can be applied. According to Ref. \cite{by}, the persistent currents are given by 
\begin{eqnarray}
\mathcal{I}=-\sum_{n,\,l}\frac{\partial E_{n,\,l}}{\partial\varsigma}.
\label{1.17}
\end{eqnarray}

Thereby, by using the eigenvalues of energy (\ref{1.15}), the persistent currents associated with the missing magnetic flux (at temperature $T=0$) are given by
\begin{eqnarray}
\mathcal{I}\vert_{T=0}=\frac{q\omega}{4\pi}-\frac{q\omega}{\pi^{3}}\mp\sum_{n}\left[\frac{q\omega}{\pi^{3}}\sqrt{1-\frac{\pi^{3}}{2\,q\,\varsigma}\left(4n+1\right)}+\frac{\omega\left(4n+1\right)}{4\varsigma\sqrt{1-\frac{\pi^{3}}{2\,q\,\varsigma}\left(4n+1\right)}}\right].
\label{1.18}
\end{eqnarray}

Hence, Eq. (\ref{1.18}) is an analogue of the persistent currents \cite{by}, i.e., it is the expression of the persistent currents associated with the missing magnetic flux $\varsigma$. Despite the influence of the topology of the screw dislocation on the energy levels (\ref{1.15}), there is no influence of the topology of the screw dislocation on the analogue of the persistent currents (\ref{1.18}). Besides, this analogue of the persistent currents exists under the restriction (\ref{1.16}), otherwise, we would have an imaginary term in the persistent currents (\ref{1.18}).

As discussed in Refs. \cite{per1,per2,per3,per4}, since the Aharonov-Bohm quantum phase \cite{ab} can influence the motion of electrically charged particles, therefore, the persistent currents is a way of observing the Aharonov-Bohm effect in mesoscopic rings. From this perspective, the persistent currents could also be a way of searching for quantum effects associated with the missing magnetic flux $\varsigma$, since the missing magnetic flux influences the interaction between the point charge and the magnetic field in the region $r\,>\,r_{0}$.

\section{Conclusions}

We have analysed the interaction of a point charge with an inhomogeneous magnetic field in an elastic medium that possesses a screw dislocation. The inhomogeneous magnetic field fills the whole space except of the inside of a region with a cylindrical shape of radius $r_{0}$. We have assumed that $r=r_{0}$ is a cut-off point, thus, we have shown that bound state solutions to the Schr\"odinger equation can be obtained. One of the characteristics of the energy levels observed is the influence of topology of the screw dislocation on them. This influence is viewed through the effective angular momentum $\gamma$ that gives rise to an Aharonov-Bohm-type effect for bound states \cite{pesk,fur,fur2,fur3,vb,vb2,vb3}. In the limit $\beta=0$, we obtain the energies levels in the absence of the screw dislocation.

Another characteristic of the energy levels is the dependence of them on the the missing magnetic flux $\varsigma$. This influence of the missing magnetic flux determines the degeneracy of the energy levels and imposes a limit on the radial quantum number. In addition, since the missing magnetic flux inside the region with a cylindrical shape of radius $r_{0}$ can be viewed as the inverse of the Aharonov-Bohm proposal \cite{magdot}, thus, the dependence of the energy levels on the missing magnetic flux $\varsigma$ can be considered as the inverse of the proposal of the Aharonov-Bohm effect for bound states \cite{pesk}, i.e., it can be considered as the Aharonov-Bohm effect for missing magnetic flux.

Finally, by exploring the dependence of the energy levels on the missing magnetic flux $\varsigma$ and by assuming that this missing magnetic flux can vary inside the region $r\,<\,r_{0}$, we have obtained an analogue of the persistent currents \cite{by} that consists of the persistent currents associated with the missing magnetic flux. In contrast to the energy levels, the analogue of the persistent currents is not influenced by the topology of the screw dislocation. However, the persistent currents associated with missing magnetic flux does exist under the restriction given in Eq. (\ref{1.16}).

\acknowledgments{The authors would like to thank CNPq for financial support.}

\end{document}